# Local temperatures of strongly-correlated quantum dots out of equilibrium

LvZhou Ye,[1] Dong Hou,[1] Xiao Zheng,[1,*] YiJing Yan,[2] and Massimiliano Di Ventra[3,†]

[1]*Synergetic Innovation Center of Quantum Information and Quantum Physics, and Hefei National Laboratory for Physical Sciences at the Microscale, University of Science and Technology of China, Hefei, Anhui 230026, China*

[2]*Collaborative Innovation Center of Chemistry for Energy Materials, and Department of Chemical Physics, University of Science and Technology of China, Hefei, Anhui 230026, China*

[3]*Department of Physics, University of California, San Diego, La Jolla, California 92093*

Probes that measure the local thermal properties of systems out of equilibrium are emerging as new tools in the study of nanoscale systems. One can then measure the temperature of a probe that is weakly coupled to a bias-driven system. By tuning the probe temperature so that the expectation value of some observable of the system is minimally perturbed, one obtains a parameter that measures its degree of local statistical excitation, and hence its local heating. However, one anticipates that different observables may lead to different temperatures and thus different local heating expectations. We propose an experimentally realizable protocol to measure such local temperatures and apply it to bias-driven quantum dots. By means of a highly accurate open quantum system approach, we show theoretically that the measured temperature is quite insensitive both to the choice of observable and to the probe-system coupling. In particular, even with observables that are distinct both physically and in their degree of locality, such as the *local* magnetic susceptibility of the quantum dot and the *global* spin-polarized current measured at the leads, the resulting local temperatures are quantitatively similar for quantum dots ranging from noninteracting to Kondo-correlated regimes, and are close to those obtained with the traditional "local equilibrium" definition.

Temperature is a thermodynamic quantity of fundamental importance in the description of systems at equilibrium. However, the extension of this concept to systems far from equilibrium is not obvious. From an operational point of view, temperature is defined as the quantity measured by a thermometer coupled to the system with which it reaches thermal equilibrium. Precisely because the thermometer plus the system reach a state of global equilibrium when coupled, the quantity that is measured by the thermometer is then attributed to the system in the limit of weak coupling and negligible heat capacity of the thermometer. This definition is no longer valid for a system out of equilibrium such as one driven by a constant bias [1]. In this case, electron-electron interactions and electron-phonon interactions are expected to induce electronic and ionic "temperatures" different from those of the same system at equilibrium [1]. The question then is: What are these temperatures and how do we measure them *directly*?

Several solutions have been proposed. For instance, Engquist and Anderson have introduced the concept of ideal potentiometer/thermometer [2]. In this case, local chemical potential and local temperature are defined by a "local equilibrium condition": the net particle and heat current flowing through the potentiometer/thermometer are set to zero [3–11]. Although such a definition is appealingly intuitive and has been used extensively in the past, *we do not have means to directly measure heat currents* (unlike particle currents for which ammeters are available) [12], and therefore its experimental realization is very limited. Other theoretical definitions include the use of approximate distribution functions [13], information compressibility [14], or generalized fluctuation-dissipation theorems [15], to name just a few (for a more complete set of definitions see, e.g., Ref. 12). All of these, however, suffer from some limitations in their experimental realizations and therefore are also of limited use.

On the experimental side, methods have been devised to determine local temperatures by monitoring properties that are sensitive to thermal fluctuations, such as the bond rupture force [16], the junction lifetime [17, 18], the mechanical stretching distance [19], the bias-driven current noise [20], and the surface-enhanced Raman intensities [21, 22]. These probes, however, provide only an *indirect* measurement of a "local temperature" leaving the original question still open.

There is, instead, an increasing body of experimental studies in which thermal probes are coupled *directly* to driven nanoscale systems with a resolution that is approaching hundreds of nanometers or less [23–30], thus making them ideal as *local thermometers*. With these thermometers an operational definition of temperature has been proposed by imposing a *minimal perturbation condition* [31], in which the temperature of the probe, $T_p$, is varied, while monitoring some observable of the system, in such a way that the expectation value of that observable is minimally perturbed. The probe temperature satisfying this condition is then a parameter attributed to the system, which characterizes its local excitations out of equilibrium. This type of definition has been used, for instance, in the study of the thermoelectric response of nanoscale systems to applied thermal gradients, leading to the prediction of temperature oscillations [10, 32]. It is an operational definition that is relatively straightforward to implement experimentally. However, it leaves open the prescription of what type of observable one should use, and whether different observables lead to quantitatively different temperatures. In addition, in the original paper [31], the thermal probe was assumed to be a bosonic system, and hence no electric current could flow between the system and the probe. This is quite a strong limitation since the experimental probes that are being developed are generally fermionic systems [23–30]. In this case, an extra constraint related to the local chemical potential of the thermal probe needs to be introduced.

In this paper, we discuss a protocol to measure the local temperature of a system out of equilibrium that is coupled to a thermal probe as schematically shown in Fig. 1. We consider a fermionic probe and allow the measurement of an observable of the system in such a way that the minimal perturbation condition be satisfied. We show that by appropriately setting the chemical potential of the probe at the beginning of the measurement, the choice of observables quite distinct both physically and in their degree of locality lead to local temperatures that are quantitatively similar even for strongly correlated systems, and very close to those obtained with the often-used local equilibrium temperature. These results lend support to the extrapolated parameter as the "temperature" of the system.

To be specific, we consider a quantum dot (QD) in contact with two leads as sketched in Fig. 1. Under a bias voltage or a thermal gradient across the two leads, the temperatures (chemical potentials) of left and right leads are $T_L$ and $T_R$ ($\mu_L$ and $\mu_R$), respectively. To determine the local temperature $T^*$ and local chemical potential $\mu^*$ of the QD, the dot is coupled to a third lead (the probe), whose chemical potential $\mu_p$ and temperature $T_p$ are tunable.

The first step of the protocol we suggest is the determination of the chemical potential of the probe. Ideally, at zero bias (equilibrium) the minimal perturbation condition [31] should yield exactly the background equilibrium temperature $T_{\rm eq}$. This can be accomplished as follows. In the presence of an applied bias voltage or a thermal gradient [33], we first determine $\mu^*$ as

$$\mu^* = \zeta_L\,\mu_L + \zeta_R\,\mu_R. \qquad (1)$$

The weight coefficients $\zeta_L$ and $\zeta_R$ are determined by [34]

$$\zeta_\alpha = 1 - \left|\frac{I_p(T_\alpha,\mu_\alpha)}{I_p(T_L,\mu_L) - I_p(T_R,\mu_R)}\right|. \qquad (2)$$

Here, $I_p(T_\alpha,\mu_\alpha)$ is the electric current measured at the probe, by setting the chemical potential and temperature of the probe to be identical with those of lead-$\alpha$.

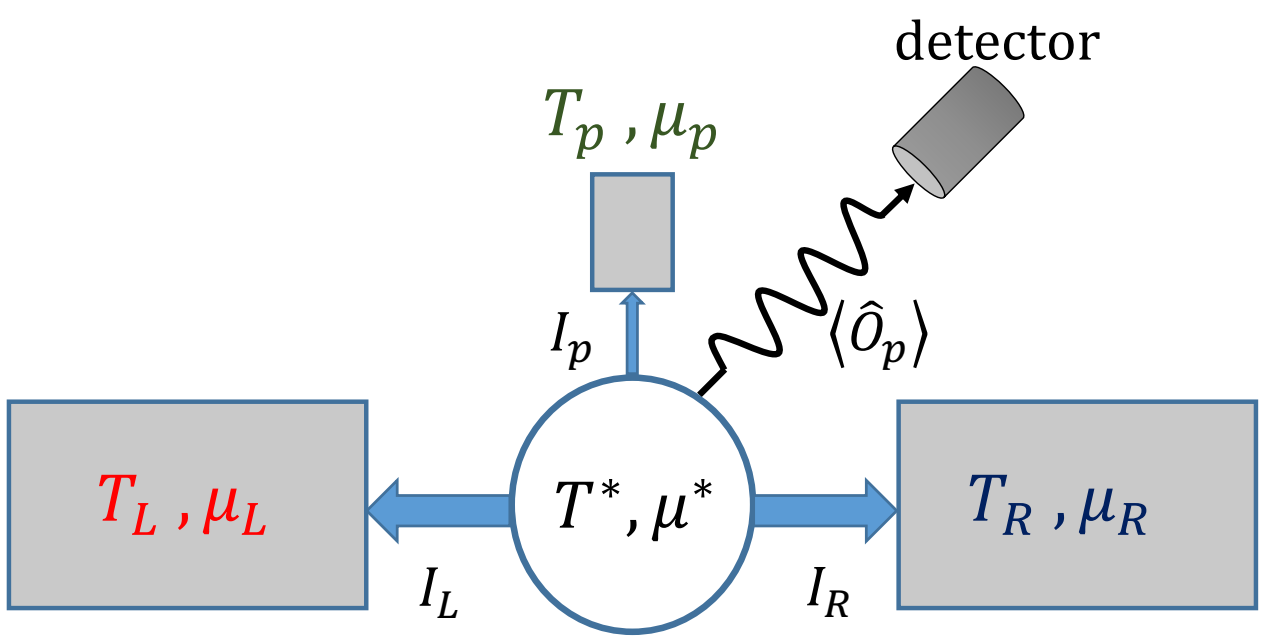

FIG. 1. Schematic diagram of the proposed protocol for the measurement of local temperature in a QD connected to two leads ($L$ and $R$). A weakly coupled probe with tunable $\mu_p$ and $T_p$ is used. The expectation value of an observable, $O_p$, is monitored while $T_p$ is varied. See the main text for details.

We then set $\mu_p = \mu^*$, and monitor the change of a given system observable $O = \langle\hat{O}\rangle$ as $T_p$ varies. The local temperature $T^*$ is finally determined by the minimal perturbation condition [31]

$$T^* = \arg\min_{T_p} \left|O_0 + \delta O_p - O_p(T_p,\mu^*)\right|. \qquad (3)$$

Here, $O_0$ is the expectation value of $\hat{O}$ measured without the probe, while $O_p(T_p,\mu^*)$ is its measured value with the probe coupled to the dot. The nonzero probe-dot coupling, $\Delta_p$, results in a finite perturbation to the intrinsic dot properties. This effect is accounted for by the correction term $\delta O_p$ in Eq. (3) determined by

$$\delta O_p = \zeta_L\,O_p(T_L,\mu_L) + \zeta_R\,O_p(T_R,\mu_R) - O_0. \qquad (4)$$

Here, $O_p(T_\alpha,\mu_\alpha)$ is measured by setting $T_p$ and $\mu_p$ to be identical to their counterparts of lead-$\alpha$.

Note that the protocol proposed above is easily realizable experimentally, and is universally applicable to any QD system. Also, for all the QDs studied in this work, the minimal perturbation of $\langle\hat{O}\rangle$ searched for in Eq. (3) is always found to be zero, *i.e.*, $O_p(T_p,\mu^*) = O_0 + \delta O_p$ can always be satisfied at a certain $T_p$. This indicates that the chosen observable $O_p$ changes monotonically and sensitively as $T_p$ varies in the vicinity of $T^*$.

It is important to verify that at exactly zero bias the local temperature measured by the minimal perturbation condition is exactly the physical equilibrium temperature $T_{\rm eq}$. This can be seen from Eqs. (1)–(4). At zero bias, we find that from Eq. (1) the local chemical potential is just the equilibrium Fermi energy. Then Eq. (3) reduces to the simple form $O_p(T_p) = O_p(T_{\rm eq})$, with the trivial solution $T_p = T_{\rm eq}$.

We now apply Eqs. (1)–(4) to QD systems described by a single-impurity Anderson model (SIAM) [35, 36]. The total Hamiltonian is $H = H_{\rm dot} + H_{\rm lead} + H_{\rm coupling}$. The dot is described by $H_{\rm dot} = \epsilon_d\,\hat{n}_d + U\hat{n}_\uparrow\hat{n}_\downarrow$. Here, $\hat{n}_d =$ $\sum_s \hat{n}_s = \sum_s \hat{a}_s^\dagger \hat{a}_s$, where $\hat{a}_s^\dagger$ ($\hat{a}_s$) creates (annihilates) a spin-$s$ electron on the dot energy level $\epsilon_d$, and $U$ is the on-dot electron-electron interaction strength. $H_{\rm lead} = \sum_{\alpha k s} \epsilon_{\alpha k} \hat{d}_{\alpha k s}^\dagger \hat{d}_{\alpha k s}$ and $H_{\rm coupling} = \sum_{\alpha k s} t_{\alpha k} \hat{a}_s^\dagger \hat{d}_{\alpha k s} +$ H.c. represent the noninteracting leads and dot-lead couplings, respectively. Here, $\hat{d}_{\alpha k s}^\dagger$ ($\hat{d}_{\alpha k s}$) creates (annihilates) an electron on the orbital $|k\rangle$ of lead-$\alpha$ ($\alpha = L, R$ or $p$); and $t_{\alpha k}$ is the coupling strength between the dot level and lead orbital $|k\rangle$.

We employ a hierarchical equations of motion (HEOM) approach to compute the reduced density matrix ($\rho$) of open fermionic systems [37, 38], so as to characterize the equilibrium and nonequilibrium properties of the SIAM [39, 40]. The effect of the leads is captured by the hybridization functions $\Delta_\alpha(\omega) \equiv \pi \sum_k |t_{\alpha k}|^2 \delta(\epsilon - \epsilon_{\alpha k})$. For numerical convenience, a Lorentzian form $\Delta_\alpha(\omega) = \Delta_\alpha W^2/[(\omega-\mu_\alpha)^2+W^2]$ is adopted, where $\Delta_\alpha$ is the coupling strength between the dot and lead-$\alpha$, and $W$ is the lead band width. Hereafter, $\Delta = \Delta_L + \Delta_R$ is taken as the unit of energy. The HEOM approach has been used to study static and dynamic Kondo effects in QDs [41–45], and it is in principle exact if all orders of hierarchy expansion were included. In practice, the numerical results converge to the exact values rapidly and uniformly with the increasing truncation level of hierarchy $N_{\rm trun}$. Once the convergence is achieved, the results are guaranteed to be quantitatively accurate [38].

We compute the system observable via $O = {\rm tr}(\rho \hat{O})$, as well as the energy distribution of electric and heat currents flowing into lead-$\alpha$ as required to determine $T^*$ with the "local equilibrium condition" [46–49]

$$j_\alpha^k(\omega) = (-1)^{k+1} \left(\frac{i}{\pi}\right) (\omega - \mu_\alpha)^k \Delta_\alpha(\omega) \times \left\{\mathcal{G}^<(\omega) + 2\, i f_\beta(\omega - \mu_\alpha)\, {\rm Im}[\mathcal{G}^r(\omega)]\right\}. \quad (5)$$

Here, we have set $e = \hbar = 1$; $k = 0$ and 1 correspond to the electric and heat currents, respectively; $f_\beta(\omega)$ is the Fermi function; and the lesser and retarded Green's functions $\mathcal{G}^<$ and $\mathcal{G}^r$ are computed from correlation functions. The global electric and heat currents flowing into lead-$\alpha$ are $I_\alpha = \int d\omega\, j_\alpha^0(\omega)$ and $J_\alpha^{\rm H} = \int d\omega\, j_\alpha^1(\omega)$, respectively. Note that the electron-phonon interactions and the phonon contribution to heat current have been neglected, since their effects are negligibly small in QDs at low temperatures [50].

We consider first a noninteracting QD ($U = 0$) under a bias voltage $\mu_R = -\mu_L = \frac{1}{2}V$. The HEOM method has the virtue that, for noninteracting systems the hierarchy terminates automatically at $N_{\rm trun} = 2$ without any approximation [37]. Figure 2(a) plots $T^*$ determined by Eqs. (1)–(4) versus the background temperature $T = T_L = T_R$. We have set the dot-lead couplings to be asymmetric with $\Delta_L = 2\Delta_R$. The weight coefficients in Eq. (1) are then $\zeta_\alpha = \Delta_\alpha/\Delta$ [34], and the local chemical potential is thus $\mu^* = \frac{1}{3}\mu_L = -\frac{1}{6}V$.

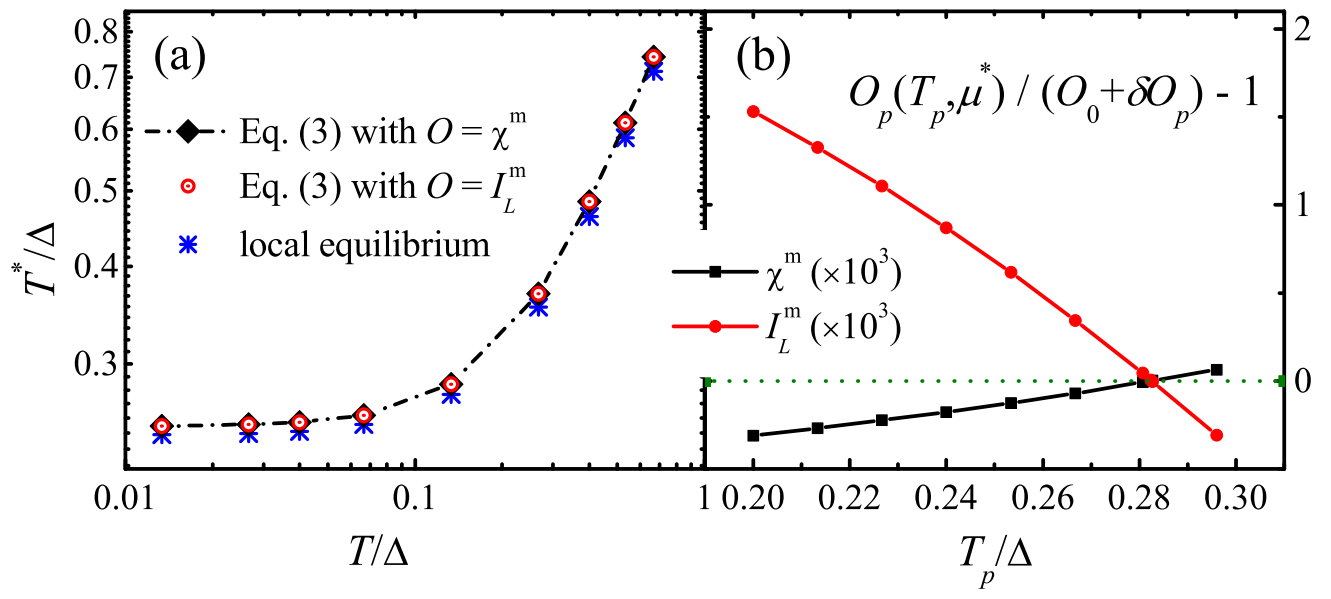

FIG. 2. (a) Local temperature $T^*$ of a noninteracting QD determined by different protocols versus background $T$. The dashed line is a guide to eyes.(b) Variation of the local magnetic susceptibility $\chi^m$ and total spin current $I_L^m$ with $T_p$ at a background $T = 0.133\Delta$. The parameters adopted are (in units of $\Delta$): $\epsilon_d = -2.67$, $U = 0$, $\Delta_L = 2\Delta_R = 0.67$, $\mu_R = -\mu_L = 0.53$, and $W = 40$.

To have an accurate measurement of $T^*$ and thus test its robustness with respect to the choice of observables, the probed observable $O$ must vary sensitively with $T_p$. We then choose some spin-related properties, because spin polarization processes require low excitation energies. These include the *local* magnetic susceptibility $\chi^m = i \int_0^t dt \langle [\hat{m}_z(t), \hat{m}_z(0)] \rangle$ with $\hat{m}_z = g\mu_B(\hat{n}_\uparrow - \hat{n}_\downarrow)/2$, and the *global* spin-polarized current $I_\alpha^m = \langle \hat{I}_{\alpha\uparrow} \rangle - \langle \hat{I}_{\alpha\downarrow} \rangle$ with $\hat{I}_{\alpha s} = i \sum_k [\hat{n}_{\alpha k s}, H_{\rm coupling}]$. Here, $g$ is the gyromagnetic ratio and $\mu_B$ is the Bohr magneton. The different degree of locality of these quantities provides an even stronger test for our protocol. As shown in Fig. 2(b), both $\chi^m$ and $I_L^m$ vary monotonically with $T_p$, and their perturbations by the probe reduce to zero at roughly the same $T_p \simeq 0.28\Delta$. Therefore, from Eq. (3) we have $T^* = 0.28\Delta$ at $T = 0.133\Delta$. Figure 2(a) compares $T^*$ obtained by different protocols over a large range of $T$. In the same plot we also show that $T^*$ determined by the minimal (zero) perturbation of $\chi^m$ or $I_L^m$ agrees closely and consistently with that obtained with the local equilibrium condition. A closer look at both the electric and heat currents that flow between the probe and the system when the minimal perturbation is satisfied reveals that they are both close to zero [51], thus explaining the agreement between the two protocols. It is also found that, under a bias voltage, $T^*$ is always higher than $T$, and their deviation increases as $T$ decreases. In particular, $T^*$ maintains a finite value ($0.25\Delta$) even as $T \to 0$. This indicates that the local heating feature becomes more visible at a lower background $T$ [1].

We now investigate interacting QDs under an antisymmetric bias voltage. We choose to examine a half-filling dot with $U = -2\epsilon_d = 2.4\Delta$. It has been shown (through the temperature-dependent conductance) that this QD exhibits prominent Kondo features at $T < T_K$ [38], with $T_K = 0.82\Delta$ being the characteristic Kondo temperature [52]. This is evident from the inset of Fig. 3(a) where the dot spectral function $A(\omega) = -\frac{1}{\pi}\text{Im}[\mathcal{G}^r(\omega)]$ is plotted for various background $T$ showing that the QD exhibits more prominent Kondo feature as $T$ lowers, as confirmed by the higher and sharper Kondo spectral peak centered at $\omega = \mu^*$. The chemical potentials of the leads are $\mu_R = -\mu_L = 0.25\Delta$. With the symmetric dot-lead couplings ($\Delta_L = \Delta_R$), the local chemical potential on the dot is $\mu^* = 0$ by Eq. (1).

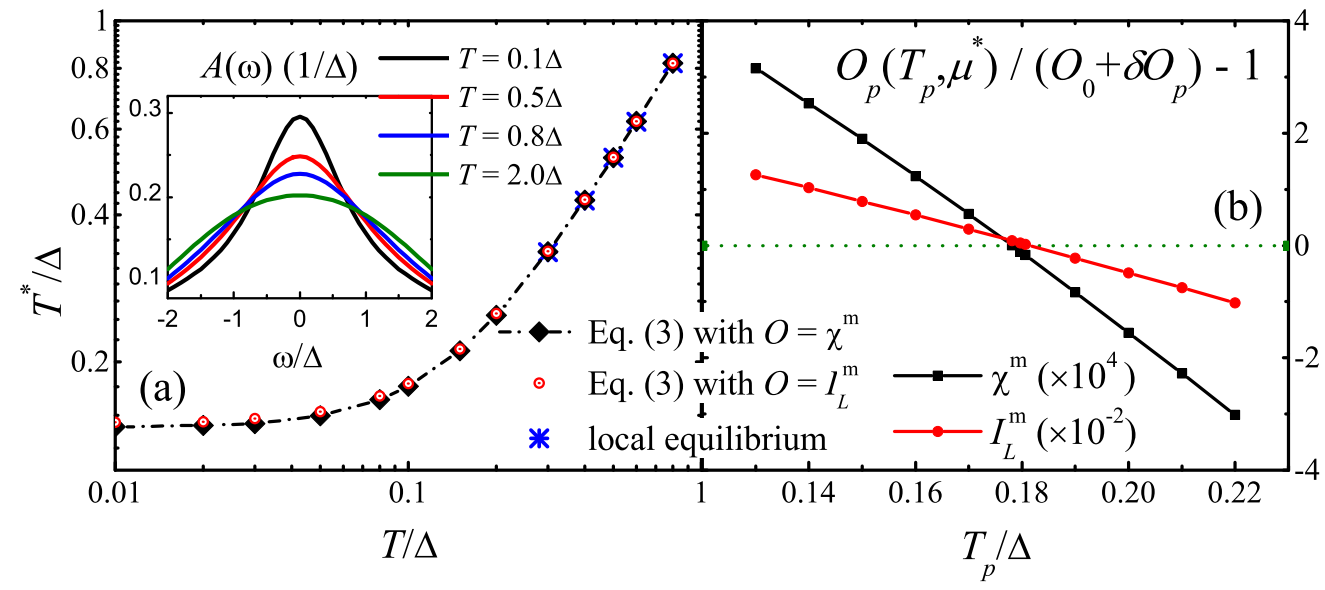

FIG. 3. (a) Local temperature $T^*$ of an interacting QD determined by different protocols versus background $T$. The dashed line is a guide to eyes. The inset shows the Kondo spectral peak of the dot at various $T$. (b) Variation of system observables $\chi^m$ and $I_L^m$ with $T_p$, at a background $T = 0.1\Delta$ lower than the Kondo temperature $T_K = 0.82\Delta$. The parameters adopted are (in units of $\Delta$): $\epsilon_d = -1.2$, $U = 2.4$, $\Delta_L = \Delta_R = 0.5$, $\mu_R = -\mu_L = 0.25$, and $W = 5$.

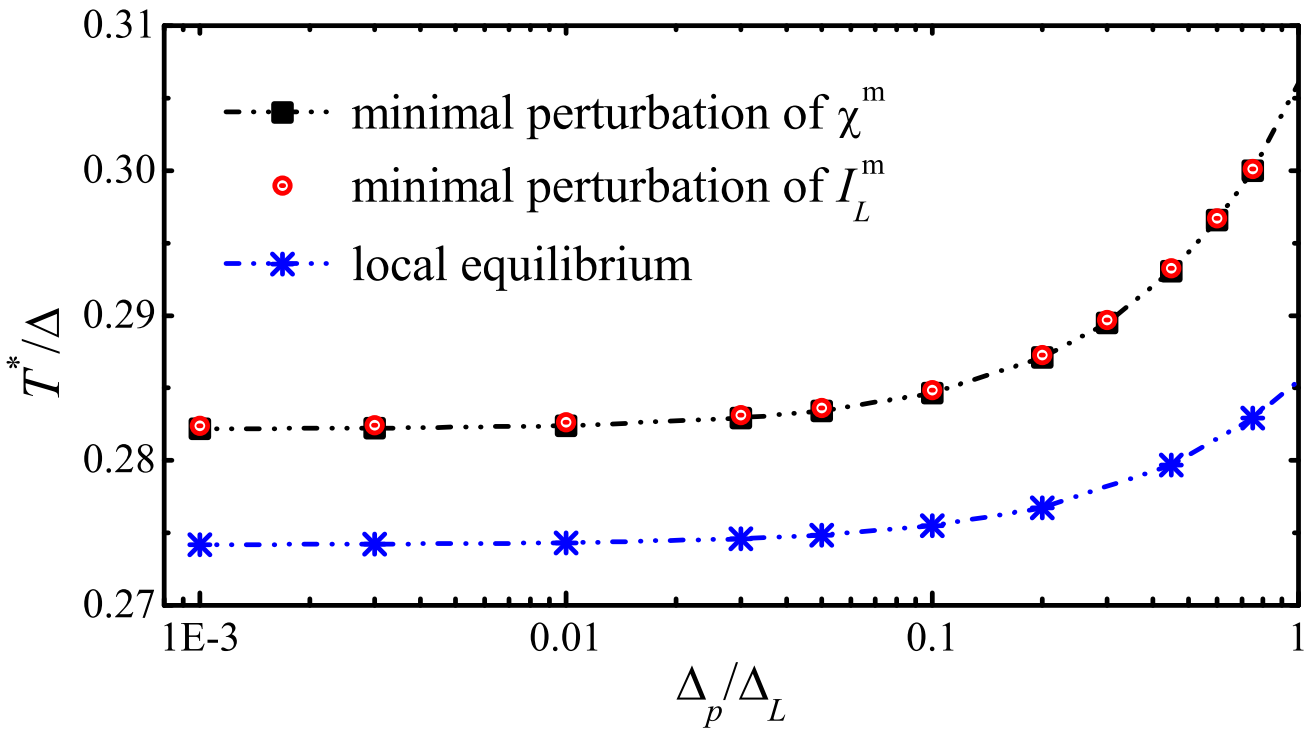

FIG. 4. Local temperature $T^*$ of a noninteracting QD determined by different protocols versus the system-probe coupling strength $\Delta_p$. The background temperature is fixed at $T = 0.133\Delta$. The parameters adopted are (in units of $\Delta$): $\epsilon_d = -2.67$, $U = 0$, $\Delta_L = 2\Delta_R = 0.67$, $\mu_R = -\mu_L = 0.53$, and $W = 40$.

Figure 3(a) compares $T^*$ determined by different protocols. Limited by computational resources at our disposal, the HEOM calculations are done at $N_{\text{trun}} = 4$ for the interacting QD [53]. The results are considered to be highly accurate within the explored range of temperatures, since the computed conductance well reproduces the Kondo scaling relation [38, 45]. As shown in Fig. 3(a), the local temperatures $T^*$ determined by the minimal (zero) perturbation of $\chi^m$ and $I_L^m$ agree remarkably well with each other, as well as with the local equilibrium temperature. This is again because the electric and heat currents flowing into the probe are both close to zero when the minimal perturbation is satisfied [51].

Similar to the noninteracting QD, the local temperature $T^*$ on the interacting dot is always higher than the background $T$, and it approaches a finite value ($T^* \simeq 0.15\Delta$) as $T \to 0$. As clearly indicated by Fig. 3(b), while both $\chi^m$ and $I_L^m$ vary monotonically with $T_p$, the latter has a much more sensitive temperature dependence. More importantly, the perturbations of both $\chi^m$ and $I_L^m$ reduce to zero at almost the same $T_p$, which highlights again the generality of our operational definition.

Finally, we investigate the influence of the coupling strength between the probe and the system. For numerical convenience we choose to examine the noninteracting QD explored in Fig. 2. Figure 4 shows $T^*$ determined by using the minimal perturbation as well as the local equilibrium conditions. As the probe-dot coupling reduces to zero, all these temperatures approach a constant value as expected. More telling though is the fact that, even with a relatively large probe-dot coupling strength (comparable with the dot-lead couplings) the resulting $T^*$ have only a minor change of $\sim 7\%$, indicating again the robustness of the proposed protocol. We thus conclude that the measured local temperature depends rather insensitively on the probe-dot coupling $\Delta_p$, which is favorable for experimental realizations.

In summary, we have proposed an operational definition of local temperature for bias-driven QDs using a "minimal perturbation condition", as represented by Eqs. (1)–(4). The same definition is also applicable to QDs subjected to external thermal gradients [54]. The operational definition applies equally well to systems ranging from noninteracting to Kondo-correlated regimes. Since this definition does not require measurements of heat currents, its experimental realization is straightforward. The "minimal perturbation condition" thus provides a useful practical means to examine local electron excitations in a nonequilibrium process, in which local heating plays an important role.

The support from the Natural Science Foundation of China (Grants No. 21233007, No. 21322305, No. 21303175, and No. 21033008), and the Strategic Priority Research Program (B) of the Chinese Academy of Sciences (XDB01020000) is gratefully appreciated. MD acknowledges support from DOE under Grant No. DE-FG02-05ER46204.

* xz58@ustc.edu.cn
† diventra@physics.ucsd.edu

**Supplemental Material for**

# Local temperatures of strongly-correlated quantum dots out of equilibrium

LvZhou Ye,[1] Dong Hou,[1] Xiao Zheng,[1,*] YiJing Yan,[2] and Massimiliano Di Ventra[3,†]

[1]*Synergetic Innovation Center of Quantum Information and Quantum Physics, and Hefei National Laboratory for Physical Sciences at the Microscale, University of Science and Technology of China, Hefei, Anhui 230026, China*

[2]*Collaborative Innovation Center of Chemistry for Energy Materials, and Department of Chemical Physics, University of Science and Technology of China, Hefei, Anhui 230026, China*

[3]*Department of Physics, University of California, San Diego, La Jolla, California 92093*

## I. REMARKS ON THE WEIGHT COEFFICIENTS $\zeta_L$ AND $\zeta_R$

Suppose $T_L = T_R$. The electric current between left and right leads is driven only by the applied bias voltage (difference in chemical potentials). With the chemical potential of probe $\mu_p$ aligned with the dot local chemical potential $\mu^*$, the net electric current flowing into the probe is exactly zero. This is because the electric currents coming from the left and right leads cancel out exactly:

$$(\mu^* - \mu_L)G_{Lp} = (\mu_R - \mu^*)G_{Rp}. \tag{S1}$$

Here, $G_{\alpha p}$ ($\alpha = L, R$) is the electron conductance between lead-$\alpha$ and the probe. Rearranging Eq. (S1), we have

$$\mu^* = \frac{G_{Rp}}{G_{Lp} + G_{Rp}}\mu_R + \frac{G_{Lp}}{G_{Lp} + G_{Rp}}\mu_L. \tag{S2}$$

By setting $\mu_p = \mu_\alpha$ and $T_p = T_\alpha$, the probe can be deemed as a part of lead-$\alpha$. Therefore, the electric current flowing through the probe $I_p$ amounts to

$$\begin{aligned} I_p(T_L, \mu_L) &= G_{Rp}\,(\mu_L - \mu_R), \\ I_p(T_R, \mu_R) &= G_{Lp}\,(\mu_R - \mu_L), \end{aligned} \tag{S3}$$

and hence $\frac{G_{Rp}}{G_{Lp}} = -\frac{I_p(T_L,\mu_L)}{I_p(T_R,\mu_R)}$. By comparing Eq. (S2) and Eq. (1) of main text, it is clear that the weight coefficient $\zeta_\alpha$ should be proportional to the two-terminal conductance $G_{\alpha p}$, as follows

$$\zeta_\alpha = \frac{G_{\alpha p}}{G_{Lp} + G_{Rp}} = 1 - \left| \frac{I_p(T_\alpha, \mu_\alpha)}{I_p(T_L, \mu_L) - I_p(T_R, \mu_R)} \right|. \tag{S4}$$

Here, we have assumed that $G_{\alpha p}$ does not change significantly with $\mu_p$ and $T_p$. Obviously, the weight coefficients should normalize to unity, $\zeta_L + \zeta_R = 1$.

For the single-impurity Anderson model (SIAM) investigated in the main text, if all leads have the same form of hybridization function, *i.e.*, $\Delta_\alpha(\omega) = \Delta_\alpha\,\eta(\omega)$ ($\alpha = L$, $R$ and $p$), we have

$$I_p(T_L,\mu_L) = \left(\frac{\Delta_p}{\Delta_L+\Delta_p}\right)\cdot\left(-\frac{2}{\pi}\right)\frac{(\Delta_L+\Delta_p)\Delta_R}{\Delta_L+\Delta_R+\Delta_p} \times \int d\omega\,\left[f_{\beta_L}(\omega-\mu_L) - f_{\beta_R}(\omega-\mu_R)\right]\,\mathrm{Im}[\mathcal{G}^r(\omega)], \tag{S5}$$

$$I_p(T_R,\mu_R) = \left(\frac{\Delta_p}{\Delta_R+\Delta_p}\right)\cdot\left(-\frac{2}{\pi}\right)\frac{(\Delta_R+\Delta_p)\Delta_L}{\Delta_L+\Delta_R+\Delta_p} \times \int d\omega\,\left[f_{\beta_R}(\omega-\mu_R) - f_{\beta_L}(\omega-\mu_L)\right]\,\mathrm{Im}[\mathcal{G}^r(\omega)], \tag{S6}$$

with $\beta_\alpha = 1/T_\alpha$. Here, the probe is deemed as part of lead-$\alpha$, and the SIAM amounts to a two-terminal model. Therefore, Eq. (9) of Ref. 1 can be used. Equations (S5) and (S6) immediately lead to the relation of

$$\frac{I_p(T_L,\mu_L)}{I_p(T_R,\mu_R)} = -\frac{\Delta_R}{\Delta_L}. \tag{S7}$$

Equation (S7) thus offers a practical means of determining the ratio $\Delta_R/\Delta_L$ in experiments. Consequently, by combining Eqs. (S4) and (S7), we have

$$\zeta_\alpha = \frac{\Delta_\alpha}{\Delta_L+\Delta_R} \tag{S8}$$

for $\alpha = L$ or $R$.

Since the electric current flowing into the probe $I_p$ can be measured straightforwardly, the weight coefficients $\{\zeta_\alpha\}$ are obtained readily from Eq. (S4). Table S1 lists the weight coefficient $\zeta_L$ for the quantum dots (QDs) studied in Figs. 2 and 3 in the main text. The numbers obtained via Eq. (S4) and Eq. (S8) agree precisely with each other.

| $\zeta_L$ | Fig. 2 noninteracting QD | Fig. 3 Kondo-correlated QD |
|---|---|---|
| Eq. (S4) | 0.6666 | 0.5000 |
| Eq. (S8) | $\frac{2}{3}$ | $\frac{1}{2}$ |

Table S1: The weight coefficient $\zeta_L$ for the QDs studied in Figs. 2 and 3 in the main text. The numbers are obtained via Eqs. (S4) and (S8), respectively.

As demonstrated clearly in the main text, the local temperature measured by using the minimal perturbation condition agrees closely and consistently with that determined by the local equilibrium condition. This is because both the electric and heat currents flowing into the probe are close to zero when the minimal perturbation condition is satisfied. Here, we verify that the values of both $I_p$ and $J_p^{\mathrm{H}}$ are indeed negligibly small when the minimal perturbation is reached. The following Tables S2 and S3 concern the noninteracting and interacting QDs investigated in the main text, respectively. It is clearly seen from the tables that, even after being scaled by the small dot-probe coupling strength $\Delta_p$, the electric and heat currents flowing through the probe are still much smaller than the currents flowing into the left and right leads.

| | scaled electric current ($e/h$) | | | scaled heat current ($\Delta/h$) | | | |
|---|---|---|---|---|---|---|---|
| $T/\Delta$ | $I_p/\Delta_p$ | $I_L/\Delta_L$ | $I_R/\Delta_R$ | $J_p^{\rm H}/\Delta_p$ | $J_L^{\rm H}/\Delta_L$ | $J_R^{\rm H}/\Delta_R$ | $(J_L^{\rm H}+J_R^{\rm H})/\Delta$ |
| 0.133 | $2.9\times10^{-3}$ | -0.37 | 0.74 | $-9.8\times10^{-3}$ | 0.16 | 0.47 | 0.26 |
| 0.667 | $-5.5\times10^{-2}$ | -0.54 | 1.08 | -0.17 | -0.25 | 1.65 | 0.39 |

Table S2: Electric and heat currents flowing into leads coupled to a noninteracting QD when the minimal (zero) perturbation condition is satisfied. The parameters adopted are (in units of $\Delta$): $\epsilon_d=-2.67$, $U=0$, $\Delta_L=2\Delta_R=100\Delta_p=0.67$, $\mu_R=-\mu_L=0.53$, and $W=40$.

| | scaled electric current ($e/h$) | | | scaled heat current ($\Delta/h$) | | | |
|---|---|---|---|---|---|---|---|
| $T/\Delta$ | $I_p/\Delta_p$ | $I_L/\Delta_L$ | $I_R/\Delta_R$ | $J_p^{\rm H}/\Delta_p$ | $J_L^{\rm H}/\Delta_L$ | $J_R^{\rm H}/\Delta_R$ | $(J_L^{\rm H}+J_R^{\rm H})/\Delta$ |
| 0.4 | $-1.3\times10^{-9}$ | -1.32 | 1.32 | $-6.9\times10^{-3}$ | 1.33 | 1.33 | 2.66 |
| 0.8 | $-4.9\times10^{-11}$ | -1.01 | 1.01 | $-3.9\times10^{-3}$ | 1.02 | 1.02 | 2.04 |

Table S3: Electric and heat currents flowing into leads coupled to an interacting QD when the minimal (zero) perturbation condition is satisfied. The parameters adopted are (in units of $\Delta$): $\epsilon_d=-1.2$, $U=2.4$, $\Delta_L=\Delta_R=100\Delta_p=0.5$, $\mu_R=-\mu_L=0.25$, and $W=5$.

## II. APPLICATION OF LOCAL EQUILIBRIUM CONDITION

Thermodynamically, the dot can be deemed as a reversed heat engine, and the leads act as heat baths as well as electron reservoirs. The internal energy of the dot is

$$\mathcal{U}=\mathcal{Q}+\mathcal{W}+\mu^*\langle\hat{n}_d\rangle, \tag{S9}$$

with $\mathcal{Q}$ and $\mathcal{W}$ being the heat and work gained by the dot, respectively. Taking the time derivatives of both sides of Eq. (S9) leads to

$$J_d^{\rm E}=J_d^{\rm H}+P+\mu^*\,\frac{d\langle\hat{n}_d\rangle}{dt}. \tag{S10}$$

Here, $J_d^{\rm E}$ ($J_d^{\rm H}$) is the energy (heat) current flowing into the dot, and $P$ is the electric power. In a stationary state, $\mathcal{U}$ is a constant, and thus $J_d^{\rm E}=d\langle\hat{n}_d\rangle/dt=0$. Therefore, $P=-J_d^{\rm H}=\sum_\alpha J_\alpha^{\rm H}$.

Figure S1(a) and (b) depict the energy distribution of bias driven electric and heat currents [$j_\alpha^0(\omega)$ and $j_\alpha^1(\omega)$] in the noninteracting QD system studied in Fig. 2 of main text, respectively. All data are computed with the hierarchical equations of motion (HEOM) approach. In the absence of the probe, $j_L^0(\omega)=-j_R^0(\omega)$ holds for any $\omega$, which ensures the conservation of electric current ($\sum_\alpha I_\alpha=0$). In contrast, the heat currents flowing through left and right leads do not cancel out ($\sum_\alpha J_\alpha^H\neq0$) – the dot behaves as a hot spot and dissipates heat into both leads.

To utilize the local equilibrium condition, the chemical potential $\mu_p$ and temperature $T_p$ of the probe are tuned until the electric and heat currents flowing into the probe [$I_p=\int d\omega\, j_p^0(\omega)$ and $J_p^{\rm H}=\int d\omega\, j_p^1(\omega)$] are both zero. Figure S1 (c) and (d) display $j_p^0(\omega)$ and $j_p^1(\omega)$ when local equilibrium is reached ($I_p=J_p^{\rm H}=0$), respectively.

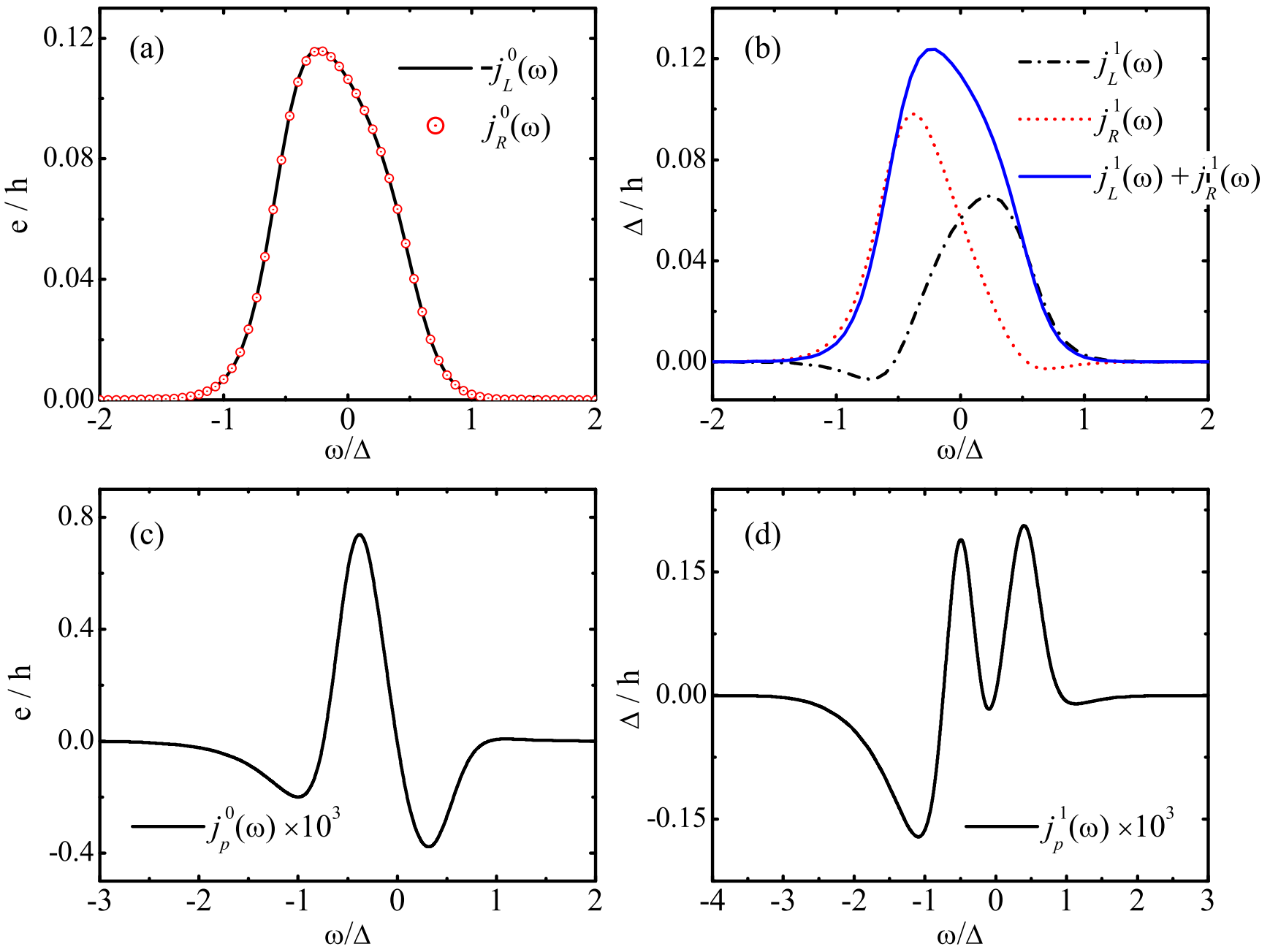

Figure S1: Panels (a) and (b) depict the energy distribution of electric and heat currents flowing into $\alpha$-lead ($\alpha = L, R$), respectively. Panels (c) and (d) plot the energy distribution of electric and heat currents flowing into the probe when the local equilibrium condition $I_p = J_p^{\mathrm{H}} = 0$ is satisfied, respectively. The parameters adopted are (in units of $\Delta$): $\epsilon_d = -2.67$, $U = 0$, $\Delta_L = 2\Delta_R = 100\Delta_p = 0.67$, $\mu_R = -\mu_L = 0.53$, $W = 40$, and $T_L = T_R = 0.133$.

## III. LOCAL TEMPERATURE OF A QUANTUM DOT SUBJECTED TO A THERMAL GRADIENT

We now demonstrate that the proposed operational definition of local temperature, Eqs. (1)–(4) of main text, is also applicable to QDs subjected to an applied thermal gradient, *i.e.*, when $\Delta T = T_R - T_L \neq 0$. Note that for this to be true, the bias voltage must be zero ($\mu_R = \mu_L$), and thus the electric current is driven only by the thermal gradient $\Delta T$.

In the absence of bias voltage, Eq. (1) of main text becomes trivial since $\mu^* = \mu_L = \mu_R$. Therefore, the chemical potential of probe is fixed at $\mu_p = \mu_L = \mu_R$, and we only need to tune $T_p$. When $T_p = T^*$, the electric current flowing into the probe vanishes. We thus have

$$(T^* - T_L)L_{Lp}^T = (T_R - T^*)L_{Rp}^T. \tag{S11}$$

Here, $L_{\alpha p}^T$ is the thermoelectric transmission coefficient between lead-$\alpha$ and probe. Through an analysis similar to that in Sec. I, one arrives at a simple expression of $T^*$ as

$$T^* = \zeta_L T_L + \zeta_R T_R. \tag{S12}$$

Here, the weight coefficients $\zeta_L$ and $\zeta_R$ are also given by the last equality of Eq. (S4). However, Eq. (S4) turns out to be a rather crude estimate of $T^*$, as shown in Fig. S2 below. The accurate measurement of $T^*$ is achieved by using Eqs. (3)–(4) of the main text.

For instance, consider a general case that dot-lead couplings are asymmetric with $\Delta_L = 5\Delta_R$. The weight coefficients $\zeta_L = \frac{5}{6}$ and $\zeta_R = \frac{1}{6}$ agree well with the "measured" values of $\zeta_L = 0.8333$ and $\zeta_R = 0.1667$ by Eq. (S4). Figure S2 compares $T^*$ obtained by different approaches. Clearly, $T^*$ determined by minimal (zero) perturbation of $\chi^m$ and $I_L^m$ agree remarkably well with the prediction of local equilibrium condition.

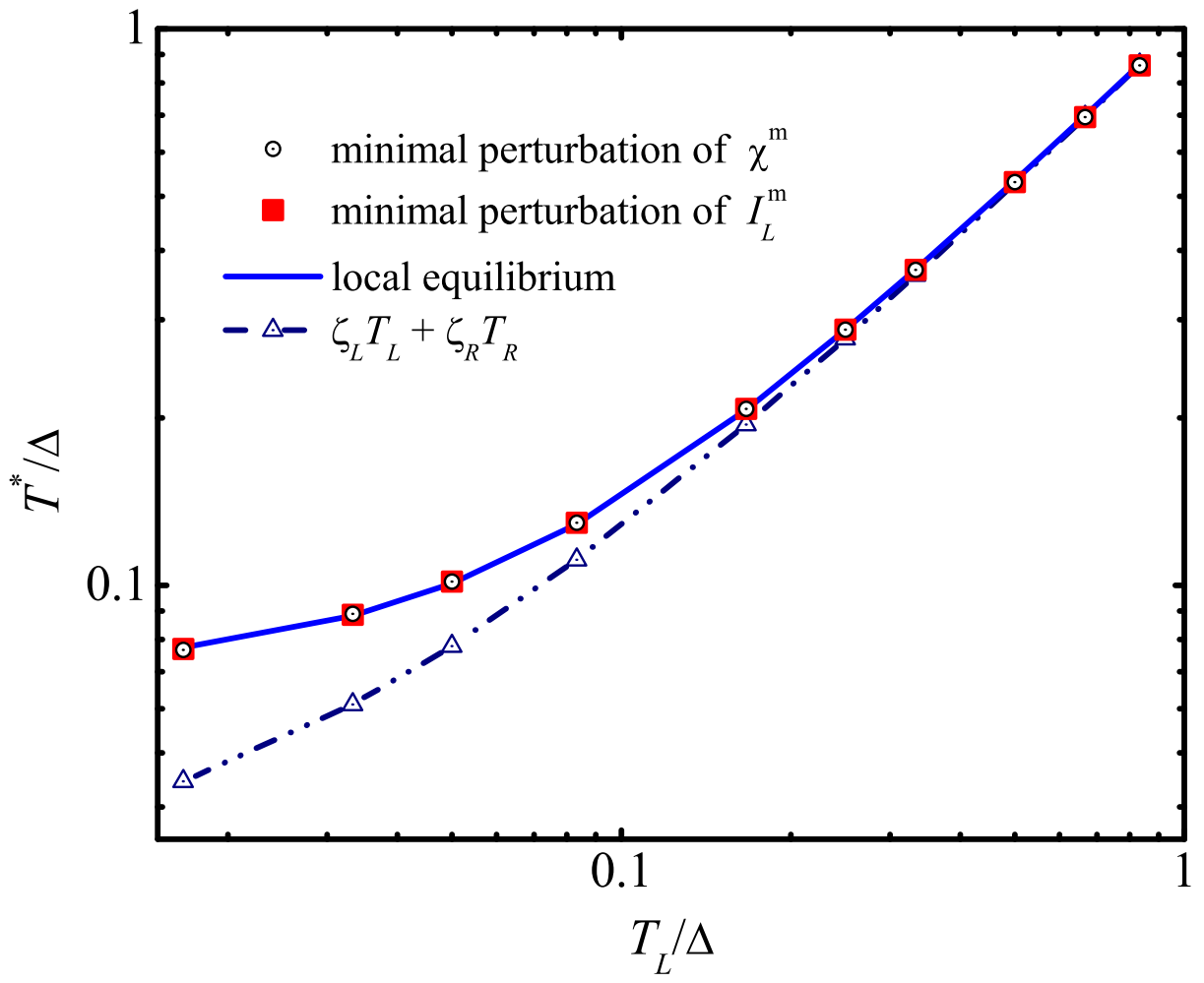

Figure S2: $T^*$ of a noninteracting QD versus the temperature of left lead $T_L$. The parameters adopted are (in units of $\Delta$): $\epsilon_d = -3.33$, $U = 0$, $\Delta_L = 5\Delta_R = 0.83$, $\mu_R = \mu_L = 0$, $\Delta T = T_R - T_L = 0.17$, and $W = 50$.

---

* Electronic address: xz58@ustc.edu.cn

† Electronic address: diventra@physics.ucsd.edu